\documentclass[A4paper,11pt]{article}
\usepackage[utf8]{inputenc}
\usepackage[longnamesfirst]{natbib}

\usepackage[top=38truemm,bottom=38truemm,left=30truemm,right=30truemm]{geometry}

\usepackage{mathtools}
\usepackage{amsthm, amssymb, amsmath, appendix}
\usepackage{url} 

\usepackage{enumerate}

\theoremstyle{definition}
\newtheorem{definition}{Definition}
\newtheorem{assumption}{Assumption}
\newtheorem{example}{Example}
\newtheorem{remark}{Remark}

\theoremstyle{theorem}
\newtheorem{theorem}{Theorem}

\newtheorem{proposition}{Proposition}
\newtheorem{corollary}{Corollary}

\usepackage{tikz}
\usetikzlibrary{intersections, calc, arrows.meta}


\title{When is it (im)possible to respect all individuals' preferences under uncertainty?\thanks{The author is grateful to two referees for comments that greatly improved the
paper. The author also thanks Yutaro Akita, Bach Dong-Xuan, Noriaki Kiguchi, Kaname Miyagishima, Nozomu Muto, Marcus Pivato, Koichi Tadenuma, Norio Takeoka, Takashi Ui, Takuro Yamashita, Tsubasa Yamashita, and Shohei Yanagita for their helpful comments and insightful discussions. This paper was presented at Online Social Choice and Welfare Seminar Series and workshops held at Hitotsubashi University and Tokyo University of Science.
This research is financially supported by KAKENHI (No.~25KJ1298).
}}
\author{Kensei Nakamura\thanks{Graduate School of Economics, Hitotsubashi University, Kunitachi, Tokyo 186-8601, Japan. E-mail: kensei.nakamura.econ@gmail.com (ORCID: 0009-0008-4549-4215)}}
\date{This version : \today}

\begin{document}
\sloppy

\maketitle

\begin{abstract}
    When aggregating Subjective Expected Utility preferences, the Pareto principle leads to an impossibility result unless the individuals have a common belief. 
    This paper examines the source of this impossibility in more detail by considering the aggregation of a general class of incomplete preferences that can represent gradual ambiguity perceptions. 
    Our result shows that the planner cannot avoid ignoring some individuals unless there is a probability distribution that all individuals agree is most plausible. 
    This means that even if individuals have similar ambiguity perceptions, the impossibility persists as long as some individual's most plausible belief differs even slightly from that of others.
    \vspace{2mm}
\end{abstract}

\vspace{5mm}

\section{Introduction}

The Pareto principle---the requirement that the social evaluation should respect unanimity among individuals---is widely accepted in economics and social choice theory. 
It is no exaggeration to say that this principle is a fundamental requirement when analyzing markets, mechanisms, and welfare criteria.

However, it is known that this principle leads to  unacceptable conclusions under uncertainty. \citet{hylland1979impossibility} and
\citet{mongin1995consistent,mongin1998paradox} showed that when aggregating Subjective Expected Utility  preferences, the Pareto principle leads to  dictatorships if there is a disagreement of beliefs. 
Following them, many papers have examined preference aggregation under uncertainty (e.g., \citet{gilboa2004utilitarian,chambers2006preference,alon2016utilitarian,danan2016robust,zuber2016harsanyi,billot2021utilitarian}). 

The difficulty of respecting unanimous agreements under uncertainty stems from tensions between social rationality (the assumption that the social preference is a Subjective Expected Utility model) and the Pareto principle. 
As is well-known, social rationality is closely related to the axiom of vNM independence (under \citeauthor{anscombe1963definition}'s (1963) framework).\footnote{\citet{chambers2006preference} examined the negative result in \citeauthor{Savage1954-SAVTFO-2}'s (1954) setup and identified which axioms lead to the impossibility.}
To clarify how severe these tensions are, \citet{zuber2016harsanyi} examined whether the social preference can satisfy the Pareto principle if we impose a much weaker independence condition, and concluded that it is difficult to respect all individuals' preferences even under these weaker assumptions. 
For a related work, see \citet{mongin2015ranking}.

Another approach to relaxing the assumption of social rationality is to drop completeness or transitivity.
\citet{danan2016robust} studied  the difficulty of the Pareto principle when aggregating Bewley preferences,  an incomplete multiprior version of the Subjective Expected Utility model (cf.~\citet{bewley2002knightian}). 
They showed that the Pareto principle implies that (i) vNM functions are aggregated by a weighted utilitarian rule à la \citet{harsanyi1955cardinal} and (ii) the planner's belief set is in the intersection of belief sets of individuals who are assigned  positive weights. 
Therefore, if the intersection of all individuals' belief sets is empty, then the planner has to ignore some individuals' preferences by assigning them a weight of 0, that is, the impossibility still holds.%
\footnote{See also \citet{pivato2024bayesian} and \citet{kurata2025reservation} for the aggregation of Bewley preferences.}

This paper provides a new impossibility theorem about  preference aggregation under uncertainty with the Pareto principle. 
We consider agents with gradual ambiguity perceptions, or more precisely, agents with variational Bewley preferences proposed by \citet{faro2015variational}. 
This class is a generalization of Bewley preferences examined by \citet{danan2016robust} and an incomplete version of variational preferences studied by \citet{maccheroni2006ambiguity}. 
In the variational Bewley model, an agent has a vNM function $u$ and a function $c$ representing their ambiguity perception, and evaluates an act $f$ to be weakly better than another act $g$ if 
\begin{equation*}
    \sum_{s\in S} p(s) u(f(s)) + c(p) \geq  \sum_{s\in S} p(s) u(g(s))
\end{equation*}
holds for each probability distribution $p$. If $c(p)$ is large, then the above inequality is easy to satisfy, so the smaller $c(p)$ is, the more plausible the agent considers the prior $p$ to be. 
By introducing the function $c$, this model captures various ambiguity perceptions. 

Our main result shows that the Pareto principle leads to impossibility again when aggregating the above general preferences, and clarifies a sufficient condition under which this principle leads to negative consequences.
Roughly speaking, the planner cannot avoid ignoring some individuals unless there is a probability distribution that all individuals unanimously consider \textit{most} plausible. 
In other words, even if individuals have similar ambiguity perceptions, the impossibility holds as long as some individual's most plausible belief differs even slightly from those of others.
Therefore, our result shows that it is difficult to satisfy the Pareto principle and avoid ignoring some individuals at the same time, even if we consider a social preference that can respect individuals' beliefs flexibly. 

This result aligns with findings about the efficiency of constant allocations in the context of speculative trade under uncertainty. 
\citet{billot2004sharing} and \citet{kajii2009interim} examined markets with agents who have multiprior preferences---Maxmin Expected Utility preferences (cf.~\citet{gilboa1989maxmin}) or Bewley preferences---under no aggregated uncertainty. 
They showed that constant allocations are Pareto optimal if and only if there is at least one probability distribution that all agents deem plausible. 
This coincides with the condition  for respecting all individuals' preferences when aggregating Bewley preferences (cf.~\citet{danan2016robust}). 
\citet{rigotti2008subjective} studied agents with a more general class of convex preferences, including variational preferences (the complete and transitive version of variational Bewley preferences). 
As a special case of their result, they showed that if individuals have variational preferences, society can reach a Pareto optimal allocation without betting if and only if the intersection of sets of priors that each individual thinks to be most plausible is nonempty. 
The condition we derive also aligns with the finding by \citet{rigotti2008subjective}, although the proofs are not related to each other. 
Thus, our result again highlights the importance of agreement about the most plausible beliefs in a distinct context.

Furthermore, our result has a technical contribution as well. 
We argue that one direction of Theorem 1 of \citet{danan2016robust} follows from our result as a corollary. 
Their result was proved using complex arguments related to convex analysis.\footnote{See the proof of Lemma 1 of \citet{danan2016robust}.} 
On the other hand, our proof is very simple since we only use the standard separating hyperplane theorem. 

Preference aggregation with gradual ambiguity perceptions was also considered in \citet{dong2024aggregation} and \citet{Mudekereza2025WP}. 
They mainly considered the decision-making model examined in \citet{cerreia2025making}, which is a special case of variational preferences (\citet{maccheroni2006ambiguity}). 
Dong-Xuan studied the case where all individuals have a common vNM function and thus focused on the aggregation of ambiguity perceptions. 
Mudekereza examined the sequential process where the planner first aggregates the subrelations induced by the individual preferences and then selects the optimal probability model (i.e., a set of priors). Since these subrelations can be represented by the Bewley model, the aggregation step is essentially based on the theorems of \citet{danan2016robust}, as Mudekereza argued.

This paper is organized as follows: Section 2 introduces the setup studied in this paper. Section 3 gives the main results and two corollaries. Section 4 discusses our result in more detail.
Section 5 provides concluding remarks.

\section{Setup}

Let $S$ be a finite set of states of the world. 
Let $N = \{ 1, 2, \cdots, n \}$ be the set of individuals with $n \geq 2$. The social planner is indexed by $0$. The set of outcomes is denoted by $X$. 
We assume that $X$ is a finite-dimensional Euclidean space. 
An act is a function $f: S \rightarrow X$. 
Let $\mathcal{F}$ be the set of  acts. 
Let $\Delta (S)$ be the set of probability distributions over $S$, that is, $\Delta (S) = \{ p\in \mathbb{R}_+^{S} ~ | ~  \sum_{s\in S} p(s)  = 1\}$.\footnote{Let $\mathbb{R}$ (resp.~$\mathbb{R}_{+}$) denote the set of real numbers (resp.~nonnegative numbers).  Let $0 \cdot (+\infty) = 0$. }

For each $i\in N\cup \{ 0 \}$, $\succsim_i$ denotes $i$'s preference relation. This paper considers preferences with the following representations proposed by \citet{faro2015variational}.

\begin{definition}[Variational Bewley Representations]
\label{def_varBew}
    A preference relation $\succsim$ is 
    a \textit{\textbf{variational Bewley preference}} if there exist
    \begin{itemize}
        \item an unbounded affine function $u: X \rightarrow \mathbb{R}$ and
        \item a convex lower semicontinuous function $c: \Delta (S) \rightarrow \mathbb{R}_+\cup \{ + \infty \} $ with $\min_{p\in \Delta(S)} c(p) = 0$
    \end{itemize} 
   such that for all $f,g\in \mathcal{F}$, 
   \begin{equation*}
            f \succsim g \iff \Big[ \sum_{s\in S} p(s) u(f(s)) + c(p) \geq  \sum_{s\in S} p(s) u(g(s)) ~~~ \forall p\in \Delta(S) ~ \Big]. 
    \end{equation*}
\end{definition}

\vspace{2mm} 

This is a generalized class of Bewley preferences studied by \citet{danan2016robust} in the context of preference aggregation.  
Variational Bewley preferences can capture various ambiguity perceptions compared to the Bewley preferences. 
To see this, we introduce the formal definition of Bewley preferences. We say that $\succsim$ is a Bewley preference if  there exist an unbounded affine function $u: X \rightarrow \mathbb{R}$ and a nonempty convex closed set $P\subseteq \Delta(S)$ such that for all  $f,g\in \mathcal{F}$, 
\begin{equation*}
            f \succsim g \iff \Big[ \sum_{s\in S} p(s) u(f(s))  \geq  \sum_{s\in S} p(s) u(g(s)) ~~~ \forall p\in P ~ \Big]. 
    \end{equation*}
If $c(\Delta(S)) \subseteq \{ 0, +\infty  \}$, then a variational Bewley preference $(u,c)$ coincides with a Bewley preference $(u, P)$ with $P = c^{-1} (0)$. 
By introducing the function $c$, variational Bewley preferences accommodate gradual ambiguity perceptions. Henceforth, we call $c$ a \textit{perception function}. 

We introduce an assumption about profiles of preferences. 
The following requires that for each individual $i\in N$, there exists a pair of outcomes relevant only for individual $i$. 

\begin{assumption}[Preference Diversity]
    For all $i \in N$, there exist $\bar{x}^i, \underline{x}^i \in  X$ such that $\bar{x}^i \succ_i \underline{x}^i$ and $\bar{x}^i \sim_j \underline{x}^i$ for all $j \in N \backslash \{ i \}$. 
\end{assumption}

As shown by \citet{weymark1993harsanyi}, Preference Diversity is equivalent to the condition that individuals' utility functions are linearly independent. 
This assumption has been imposed in several papers (e.g., \citet{mongin1998paradox,danan2016robust,zuber2016harsanyi}).

\begin{remark}[Properties of variational Bewley preferences]
\label{rem_pro}
    To facilitate better understanding, we make several remarks on variational Bewley preferences  here. 
    \begin{enumerate}[(i)]
        \item Variational Bewley preferences may violate both completeness and transitivity but are complete and transitive over constant acts.
        \item They satisfy statewise dominance (note that to satisfy this property, the condition that $\min_{p\in \Delta(S)} c(p) = 0$ in Definition \ref{def_varBew} is necessary).  
        \item They may violate independence but satisfy one direction of it (i.e., for all $f, g, h\in \mathcal{F}$ and $\alpha \in (0,1)$, $f\succsim g$ implies $\alpha f + (1 - \alpha )h\succsim \alpha g + (1 - \alpha )h$). 
        \item They satisfy transitivity if and only if they satisfy independence if and only if they are Bewley preferences. 
    \end{enumerate}
\end{remark}

    

\begin{remark}[Unboundedness]
    Note that in the standard definition of Bewley preferences, $u$ is not necessarily unbounded (cf.~\citet{gilboa2010objective,danan2016robust}). 
    We introduce this condition only to make the comparison with the variational Bewley preferences clearer. 
    For a more detailed discussion, see Remark  \ref{remark:final}. 
\end{remark}

\section{Results}

We consider the standard Pareto principle, which is widely accepted as a property of aggregation. 

\begin{description}
    \item[\bf Standard Pareto.] For all $f,g \in \mathcal{F}$, if $f\succsim_i g$ for all $i\in N$, then $f\succsim_0 g$.
\end{description}

The following is our main result. 
It offers a necessary condition to satisfy Standard Pareto when aggregating variational Bewley preferences. 

\begin{theorem}
\label{thm:main}
 Suppose that for each $i\in N\cup \{ 0 \}$, $\succsim_i$ is  a variational Bewley preference $(u_i, c_i)$ and  Preference Diversity holds. 
Then, Standard Pareto implies that there exists $(\alpha, \beta) \in \mathbb{R}^n_+ \times \mathbb{R}$   such that 
    \begin{equation*}
        u_0 = \sum_{i\in N} \alpha_i u_i + \beta, ~~ \text{and} ~~
    c_0 (p) \geq \max_{i\in N} \alpha_i c_i(p) ~~~ \forall p\in \Delta(S).
    \end{equation*}
\end{theorem}

Before providing a proof, we discuss the implications of this result. 
This theorem states that vNM functions are aggregated in a utilitarian way, as in \citeauthor{harsanyi1955cardinal}'s (\citeyear{harsanyi1955cardinal}) theorem. 
Furthermore, the planner's perception function is bounded below using the individuals' perception functions and the weight vector used to aggregate the vNM functions. 
To see this condition in more detail, we consider a result immediately following from Theorem \ref{thm:main}. 



\begin{corollary}
\label{cor:impos}
Suppose that for each $i\in N\cup \{ 0 \}$, $\succsim_i$ is  a variational Bewley preference $(u_i, c_i)$ and  Preference Diversity and Standard Pareto hold. 
Then, $ u_0 = \sum_{i\in N} \alpha_i u_i + \beta$ for some $(\alpha, \beta) \in \mathbb{R}^n_+ \times \mathbb{R}$,  and
\begin{equation*}
    c^{-1}_0 (0) \subseteq \bigcap_{i\in N : \alpha_i > 0} c^{-1}_i (0). 
\end{equation*} 
\end{corollary}

Note that by the definition of variational Bewley preferences, $c^{-1}_0 (0)$ must be nonempty.
Therefore, if  $\bigcap_{i\in N} c_i^{-1} (0) = \emptyset $, then it follows from Corollary \ref{cor:impos} that for at least one individual $i$, $\alpha_i = 0$ must hold---that is, $i$'s preference is ignored by the social planner.  
This observation implies that to assign all individuals positive weights, the intersection $\bigcap_{i\in N} c_i^{-1} (0)$ of most plausible priors must be nonempty. 

Furthermore, we can derive one direction of Theorem 1 of \citet{danan2016robust} as an immediate result of Corollary \ref{cor:impos} (note that the other direction also holds in this case and follows from a direct calculation). 
The next result can be obtained from the observation that  if a variational Bewley preference $(u_i, P_i)$ is a Bewley preference $(u_i , P_i)$, then $P_i = c_i^{-1} (0) $.\footnote{Precisely speaking, Corollary \ref{cor_dan} is slightly different from the corresponding part of Theorem 1 of \citet{danan2016robust}: While they only assumed that $X$ is a convex subset of some Euclidean space and  $u_i$ is nonconstant for each $i\in N\cup\{0 \}$, we consider the case where $X$ is a finite-dimensional Euclidean space and $u_i$ is unbounded. However, this difference is not essential. For a detailed explanation, see Remark \ref{remark:final}.} 

\begin{corollary}
\label{cor_dan}
Suppose that for each $i\in N\cup \{ 0 \}$, $\succsim_i$ is  a Bewley preference $(u_i, P_i)$ and  Preference Diversity and Standard Pareto holds. 
Then,  there exists $(\alpha, \beta) \in \mathbb{R}^n_+ \times \mathbb{R}$   such that 
    \begin{equation*}
        u_0 = \sum_{i\in N} \alpha_i u_i + \beta, ~~ \text{and} ~~
    P_0 \subseteq \bigcap_{i\in N : \alpha_i > 0 } P_i. 
    \end{equation*}
\end{corollary}

Theorem 1 of \citet{danan2016robust} clarified when the social planner cannot respect all individuals' preferences under Standard Pareto: Some individuals have to be "excluded" if $\bigcap_{i\in N} P_i = \emptyset$. 
In other words, this result claims that unless all individuals partially agree on their predictions, the social planner cannot assign positive weights to all of them. 

Compared to this result, our results show that the difficulty of Standard Pareto is more severe than expected. 
 Corollary \ref{cor:impos} states that if the intersection of sets of probability distributions each individual thinks to be most plausible is empty, then the planner cannot avoid excluding some individuals' preferences. 
This means that even if the individuals have perception functions that are very similar to but slightly different from one another, the planner cannot respect all individuals' preferences and, moreover, has to evaluate all alternatives based on only one individual's preference. 

We then provide a proof of Theorem \ref{thm:main}.  
The proof is simple: We only use the separating hyperplane theorem. 
Thus, this paper can be considered as providing a simple alternative proof of one direction of Theorem 1 of \citet{danan2016robust}. 

\begin{proof}[\bf Proof of Theorem \ref{thm:main}.]
\textit{Aggregating utility functions.} 
For all $x,y \in X$,  $x\succsim_i y$ for all $i\in N$ implies that $x \succsim_0 y$. 
By the result of \citet{de1995note}, there exists  $(\alpha, \beta) \in \mathbb{R}^n_+ \times \mathbb{R}$  such that $u_0 = \sum_{i\in N} \alpha_i u_i + \beta$. If $\alpha = 0$, then  $u_0$ is a constant function, which is a contradiction since $u_0$ is nonconstant. 

\begin{figure}
    \centering
    \moveleft 2mm \vbox{\includegraphics[width=1.0\textwidth]{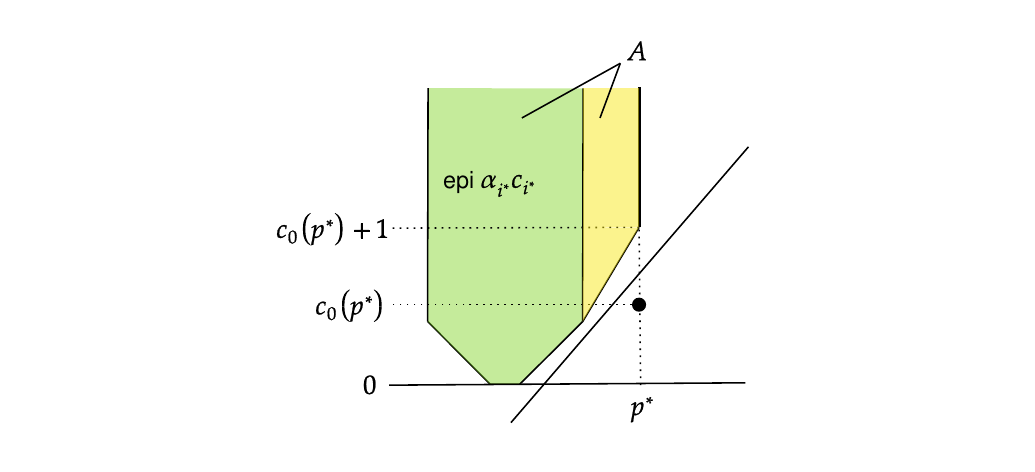}}
    \caption{The set $A$ and a separating hyperplane}
    \label{fig:1}
\end{figure}

\textit{Aggregating perception functions.} 
Suppose to the contrary that there exists $p^\ast \in \Delta (S)$ such that $c_0 (p^\ast) < \max_{i\in N} \alpha_i c_{i}(p^\ast)$. 
Then we fix $i^\ast \in N$ such that  $c_0(p^\ast) < \alpha_{i^\ast} c_{i^\ast} (p^\ast)$.
The set $\text{epi} \, \alpha_{i^\ast} c_{i^\ast}  \coloneqq  \{ (p,\gamma) \in \Delta (S) \times \mathbb{R}_+  ~ | ~ c_{i^\ast} (p) < + \infty  ~~ \text{and}~~ \gamma \geq \alpha_{i^\ast} c_{i^\ast} (p) \}$ is a convex closed  set since $c_{i^\ast}$ is a convex lower semicontinuous function,  and $(p^\ast, c_0 (p^\ast)) \notin \text{epi} \, \alpha_{i^\ast} c_{i^\ast}$. 
Define the set $A$ as the closed convex hull of $\{ (p^\ast, c_0(p^\ast) + 1 + \gamma) \mid \gamma\in \mathbb{R}_+\} \cup \text{epi} \, \alpha_{i^\ast} c_{i^\ast}$. 
Then $A$ is also a convex closed  set and $(p^\ast, c_0 (p^\ast)) \notin A$ 
(cf.~Figure \ref{fig:1}). 
By the separating hyperplane theorem, there exist $v \in \mathbb{R}^S$ and $\kappa, \lambda \in \mathbb{R}$ such that for all $p$ in the convex hull of $\{p^\ast\} \cup \{ p'\in \Delta(S) \mid c_{i^\ast} (p') < + \infty \}$, 
\begin{align}
\label{eq:fisrtsep}
    &\sum_{s\in S} p(s) v_s + \kappa \alpha_{i^{\ast }} c_{i^\ast}(p)\geq \lambda > \sum_{s\in S} p^\ast (s) v_s +  \kappa c_0 (p^\ast). 
\end{align}
If $\kappa \leq 0$, then by setting $p = p^\ast$, \eqref{eq:fisrtsep} can be rewritten as 
\begin{align*}
    &\sum_{s\in S} p^\ast (s) v_s + \kappa \alpha_{i^{\ast }} c_{i^\ast}(p^\ast) > \sum_{s\in S} p^\ast (s) v_s +  \kappa c_0 (p^\ast), 
\end{align*}
which is a contradiction to $\kappa \alpha_{i^{\ast }} c_{i^\ast}(p^\ast) \leq \kappa c_0 (p^\ast) $.\footnote{The author thanks Yutaro Akita for pointing out an error about the case where $\kappa = 0$ in the previous version.}
Therefore, $\kappa > 0$. 
By multiplying each element by  $1 / \kappa$, 
 we can assume $\kappa = 1$ without loss of generality. 
Then, for all $p \in \Delta(S)$,  
\begin{align*}
    \sum_{s\in S} p(s) v_s + \alpha_{i^{\ast }} c_{i^\ast}(p)\geq \lambda > \sum_{s\in S} p^\ast (s) v_s +  c_0 (p^\ast). 
\end{align*}
This can be rewritten as for all $p \in \Delta (S)$, 
\begin{equation}
\label{eq_var_impsep}
    \sum_{s\in S} p (s) v_s + \alpha_{i^\ast} c_{i^\ast}(p)\geq \sum_{s\in S} p (s) \lambda =  \sum_{s\in S} p^\ast (s) \lambda > \sum_{s\in S} p^\ast (s) v_s +  c_0 (p^\ast). 
\end{equation}


By Preference Diversity, there exist $\bar{x}, \underline{x} \in X$ such that $\bar{x} \succ_{i^\ast} \underline{x}$ and $\bar{x} \sim_j \underline{x}$ for all $j \in N \backslash \{ i^\ast \}$.
We can take $x$ and $f(s)$ (for each $s\in S$) on the line passing through 
 $\bar{x}$ and  $\underline{x}$ so that $\lambda = \alpha_{i^\ast} u_{i^\ast} (x) $ and $v_s =  \alpha_{i^\ast} u_{i^\ast} (f(s)) $.
By \eqref{eq_var_impsep}, for all $p\in \Delta (S)$, 
\begin{align*}
        &\sum_{s\in S} p (s) \alpha_{i^\ast}  u_{i^\ast} (f(s))  + \alpha_{i^\ast} c_{i^\ast}(p) \geq \sum_{s\in S} p (s) \alpha_{i^\ast} u_{i^\ast} (x)\\ &
        \iff 
        \sum_{s\in S} p (s)  u_{i^\ast} (f(s))  +  c_{i^\ast}(p) \geq \sum_{s\in S} p (s)  u_{i^\ast} (x), 
\end{align*}
that is, $f\succsim_{i^\ast} x$. 
By construction, $f \sim_j x$ for all $j\in N\backslash \{ i^\ast \}$. Therefore, $f \succsim_i x$ for all $i\in N$. 


Also, by \eqref{eq_var_impsep}, 
\begin{equation}
\label{eq:P1}
     \sum_{s\in S} p^\ast (s) \alpha_{i^\ast}  u_{i^\ast} (x) > \sum_{s\in S} p^\ast (s)  \alpha_{i^\ast}  u_{i^\ast} (f(s)) +  c_0 (p^\ast). 
\end{equation}
By the definitions of $x$ and $f$, 
\begin{equation}
\label{eq:P2}
    \sum_{j\in N\backslash \{ i^\ast \}} \sum_{s\in S} p^\ast (s) \alpha_j u_j (x) + \beta   = \sum_{j\in N\backslash \{ i^\ast \}} \sum_{s\in S} p^\ast (s)  \alpha_j u_j (f(s)) + \beta. 
\end{equation}
By \eqref{eq:P1} and \eqref{eq:P2}, 
\begin{align*}
    &\sum_{i\in N} \sum_{s\in S} p^\ast (s) \alpha_i u_i (x) + \beta > \sum_{i\in N} \sum_{s\in S} p^\ast (s)  \alpha_i u_i (f(s))+ \beta +  c_0 (p^\ast)\\
    &\iff \sum_{s\in S} p^\ast (s)  u_0 (x) >  \sum_{s\in S} p^\ast (s) u_0 (f(s))+  c_0 (p^\ast). 
\end{align*}
This means $f \not\succsim_0 x$, which is a contradiction to Standard Pareto. 
\end{proof}

\begin{remark}
    \label{remark:final}
     Here, we provide a technical remark about the difference between Corollary \ref{cor_dan} and Theorem 1 of \citet{danan2016robust}. 
    To directly apply our theorem to the proof of Corollary \ref{cor_dan}, we need to assume that $X$ is the entire set of some Euclidean space and that vNM functions are unbounded, because these assumptions are necessary when taking two acts $x$ and $f$ in the proof of our theorem. 
    However, when we directly prove Theorem 1 of \citet{danan2016robust} using the argument in our proof, these assumptions are not needed: 
    If $c_i$ takes only $0$ or $+\infty$ (i.e., $\succsim_i$ is a Bewley preference) for each $i\in N\cup\{ 0 \}$, then \eqref{eq_var_impsep} can be rewritten as for all $p\in c_{i^\ast}^{-1} (0)$, 
    \begin{equation}
    \label{eq:bew_sep}
    \sum_{s\in S} p (s) v_s \geq \sum_{s\in S} p (s) \lambda =  \sum_{s\in S} p^\ast (s) \lambda > \sum_{s\in S} p^\ast (s) v_s. 
    \end{equation}
    In this case, we can without loss of generality assume that $v \in [u_{i^\ast}(\underline{x}), u_{i^\ast} (\bar{x} ) ]^S$ and $\lambda \in [u_{i^\ast}(\underline{x}), u_{i^\ast} (\bar{x} ) ]$.\footnote{If a pair $(v,\lambda) \in\mathbb{R}^S\times \mathbb{R}$ satisfies \eqref{eq:bew_sep},  then $(\alpha v + \beta, \alpha\lambda+\beta)$ also satisfies \eqref{eq:bew_sep} for each $(\alpha, \beta) \in \mathbb{R}_{++} \times \mathbb{R}$. 
    Thus, by letting $\alpha$ be small enough and $\beta = {1\over 2} (u_{i^\ast}(\underline{x}) + u_{i^\ast} (\bar{x} ) )$, 
    we can construct $\alpha v + \beta \in [u_{i^\ast}(\underline{x}), u_{i^\ast} (\bar{x} ) ]^S$ and $\alpha \lambda + \beta \in [u_{i^\ast}(\underline{x}), u_{i^\ast} (\bar{x} ) ]$ that satisfy \eqref{eq:bew_sep}. }
    Hence, the proof can be applied even when $u$ is not  unbounded and $X$ is a proper convex subset of some Euclidean space. 
\end{remark}

\section{Discussion}

Finally, we discuss the violations of transitivity and independence in the social preference and the converse of Theorem \ref{thm:main}. 

\subsection{Violation of Transitivity and Independence}

As noted in Remark \ref{rem_pro}, variational Bewley preferences may violate transitivity and independence. 
While considering such individual preferences may be appropriate for describing actual individual choice patterns, it is more controversial whether it is appropriate to use such preferences for social evaluation.%
\footnote{Notice that the literature has also considered social preferences that may violate transitivity and independence. 
In the context of Arrovian preference aggregation, intransitive social preferences have been examined (e.g., \citet{sen1969quasi,mas1972general,blau1977social,blair1982acyclic,brandl2020arrovian}). For independence, since \citeauthor{diamond}'s (\citeyear{diamond}) critique of \citet{harsanyi1955cardinal}, many papers have studied social preferences that are not assumed to satisfy independence (e.g., \citet{alon2016utilitarian,hayashi2019fair,mongin2021rawls}). }
We argue that to account for gradual ambiguity perceptions, studying preferences that violate these axioms would be one of the reasonable ways. 
Starting from the Subjective Expected Utility model, the Bewley model can be characterized by the violation of completeness. 
Apart from transitivity and independence, the remaining axioms in the Subjective Expected Utility model---nontriviality, monotonicity, and continuity---are either very basic or merely technical. 
Therefore, to accommodate gradual ambiguity perceptions, it is necessary to consider preferences that may violate transitivity or independence.
Once the above argument is accepted, dropping one of these two is insufficient: As discussed in Remark \ref{rem_pro}, a variational Bewley preference that satisfies transitivity or independence is a Bewley preference. 



On the other hand, of course, we can take the position that the social preference should satisfy transitivity and independence. Since the social preference becomes a Bewley preference in this case, the following result can be obtained from our main result: 

\begin{proposition}
\label{prop:tran}
 Suppose that for each $i\in N$, $\succsim_i$ is  a variational Bewley preference $(u_i, c_i)$, $\succsim_0$ is a Bewley preference $(u_0, P_0)$, and Preference Diversity holds. 
Then, Standard Pareto holds if and only if there exists $(\alpha, \beta) \in \mathbb{R}^n_+ \times \mathbb{R}$   such that 
    \begin{equation}
    \label{eq:char_tra}
        u_0 = \sum_{i\in N} \alpha_i u_i + \beta, ~~ \text{and} ~~
    P_0 \subseteq \bigcap_{i\in N: \alpha_i > 0} c^{-1}_i (0).
    \end{equation}
\end{proposition}

\begin{proof}
  We only prove the if part since the only-if part directly follows from Theorem \ref{thm:main}. Let $f,g\in  \mathcal{F}$ be such that $f\succsim_i g$ for all $i\in N$. 
Then, for all $i\in N$ and $p\in c^{-1}_i (0)$, 
we have $\sum_{s\in S} p(s) u_i (f(s))  \geq \sum_{s\in S} p(s) u_i (g(s)) $. Therefore, by \eqref{eq:char_tra}, for all $p\in P_0$, 
\begin{equation*}
  \sum_{s\in S} p(s) u_0 (f(s)) = \sum_{i\in N} \alpha_i \sum_{s\in S} p(s) u_i (f(s)) +\beta  \geq \sum_{i\in N}\alpha_i \sum_{s\in S} p(s) u_i (g(s))+ \beta = \sum_{s\in S} p(s) u_0 (g(s)), 
\end{equation*}
that is, Standard Pareto holds. 
\end{proof}

This proposition shows that if the planner's preference satisfies transitivity and/or independence, then Standard Pareto implies that the planner's multi-prior is in the set of probability distributions that all individuals with positive weights consider most plausible. Note that this result is an if-and-only-if statement, as discussed later. 


\subsection{The Converse of Theorem \ref{thm:main}}

Recall that Theorem \ref{thm:main} is not an if-and-only-if statement because the converse does not hold. To see this, consider a two-person, two-state example.

\begin{example}
    Let $N = \{ 1,2 \}$ and $S=\{ s_0, s_1\}$. In this case,  each probability distribution over $S$ can be identified with a probability $p_1$ of $s_1$. Consider two acts $f$ and $g$ and the following preference profile: $u_1 (f(s_0)) = u_1 (g(s_0)) = u_2 (f(s_0)) = u_2 (g(s_0)) = 0$, $u_1(f(s_1)) = u_2(f(s_1)) = 2$, $u_1(g(s_1)) = u_2(g(s_1)) = 4$, and $c_1 (p_1) = c_2 (p_1) = 2p_1$ for each $p_1\in [0,1]$. Then, $f\sim_1 g$ and $f\sim_2 g$ hold. Consider the aggregation rule such that $u_0 = {1\over 2} u_1 + {1\over 2} u_2$ and $c_0 = {1\over2}c_1$. Then, this social preference violates Standard Pareto although it satisfies the conditions in Theorem \ref{thm:main}.\footnote{Note that we can construct such a preference profile without violating Preference Diversity. For instance, suppose that $X = \mathbb{R}^3$ and that $u_1 (x) = x_1 + x_3$ and $u_2 (x) = x_2 + x_3$ for all $x\in X$. Then, Preference Diversity holds, and we can take $f,g\in F$ satisfying the desired conditions by choosing $f(s)$ and $g(s)$ from $\{ (0,0,x_3)\mid x_3\in \mathbb{R} \}$ for each $s\in S$. } 
    Indeed, if $p_1 = 1$, then $u_0(f(s_1)) + c_0(1) = {1\over 2}u_1(f(s_1)) + {1\over 2} c_1(1) + {1\over 2}u_2(f(s_1)) = 3 < 4 =  {1\over 2}u_1(g(s_1)) +  {1\over 2}u_2(g(s_1)) = u_0(g(s_1))$, which implies $f\not\succsim_0 g$. 
\end{example}


Thus, our result provides only weak conditions obtained from Standard Pareto. 
Since we understand Theorem \ref{thm:main} as an impossibility result, this problem is not so severe. 
However, it is still important to examine when the converse holds, because this helps clarify the conditions under which the social planner's perception function can be constructed without violating the imposed axiom.
To address this issue, we here discuss (i) sufficient conditions under which the converse holds and (ii) an axiom that characterizes the aggregation rules obtained in Theorem \ref{thm:main}. 

First, we consider issue (i). 
We can show that if the preferences of all individuals or the social planner's preferences are assumed to be Bewley preferences, then Theorem \ref{thm:main} becomes an if-and-only-if result. 
In other words, to satisfy Standard Pareto, it is sufficient that either the planner or the individuals have preferences with transitivity and independence (cf.~Remark \ref{rem_pro} (iv)). 
For the case where the planner's preference is a Bewley preference, see Proposition \ref{prop:tran}. 
The following result considers the case where individual preferences are Bewley preferences. 

\begin{proposition}
\label{prop:iff_ind}
Suppose that for each $i\in N$, $\succsim_i$ is a  Bewley preference $(u_i, P_i)$, $\succsim_0$ is a variational Bewley preference $(u_0, c_0 )$, and Preference Diversity holds. 
Then, Standard Pareto holds if and only if there exists $(\alpha, \beta) \in \mathbb{R}^n_+ \times \mathbb{R}$   such that 
    \begin{equation}
    \label{eq:char_tra_ind}
        u_0 = \sum_{i\in N} \alpha_i u_i + \beta, ~~ \text{and} ~~
        \mathrm{dom}\,  c_0  \subseteq \bigcap_{i\in N: \alpha_i > 0} P_i, 
    \end{equation}
    where $\mathrm{dom}\,  c_0 = \{ p\in\Delta (S) \mid c_0 (p) < + \infty  \}$. 
\end{proposition}

\begin{proof}
We only prove the if part since the only-if part directly follows from Theorem \ref{thm:main}. 
Let $f,g\in  \mathcal{F}$ be such that $f\succsim_i g$ for all $i\in N$. 
Then, for all $i\in N$, 
we have $\sum_{s\in S} p(s) u_i (f(s))  \geq \sum_{s\in S} p(s) u_i (g(s)) $ for all $p\in P_i$. Therefore, by \eqref{eq:char_tra_ind} and $c_0 \geq 0$, for all $p\in \Delta (S)$, 
\begin{align*}
  \sum_{s\in S} p(s) u_0 (f(s)) + c_0 (p)&= \sum_{i\in N} \alpha_i \sum_{s\in S} p(s) u_i (f(s)) + \beta  + c_0 (p) 
  \geq \sum_{i\in N} \alpha_i  \sum_{s\in S} p(s) u_i (g(s)) + \beta 
  \\& = \sum_{s\in S} p(s) u_0 (g(s)), 
\end{align*}
that is, Standard Pareto holds. 
\end{proof}

This proposition shows that if individuals have Bewley preferences, then any perception function that satisfies the condition \eqref{eq:char_tra_ind} is compatible with Standard Pareto. 

Next, we address issue (ii), that is, axioms that characterize the aggregation rules in Theorem \ref{thm:main}. For $i\in N$, let $\mathcal{F}^i \subseteq \mathcal{F}$ be such that if $f\in \mathcal{F}^i$, then for each $s\in S$, $f(s)$ is on the line passing through  $\bar{x}^i$ and $\underline{x}^i$, where these two are the elements in $X$ introduced in the definition of Preference Diversity. 
For all $f,g\in \mathcal F^i$, by the definition of $\bar{x}^i$ and $\underline{x}^i$, we have $f\sim_j g$ for all $j\in N\setminus \{ i \}$. 
Therefore, $\mathcal{F}^i$ is the set of acts that is relevant only for individual $i$. 

We then consider three axioms weaker than Standard Pareto. 

\begin{description}
  \item[\bf Constant Pareto.] For all $x, y\in X$, if $x \succsim_i y$ for all $i\in N$, then $x\succsim_0 y$. 
\end{description}

\begin{description}
  \item[\bf Liberalism.] For all $i\in N$ and $f,g\in \mathcal{F}^i$, if $f\succsim_i g$, then $f\succsim_0 g$. 
\end{description}

\begin{description}
  \item[\bf Dominance Pareto.] For all $f,g\in \mathcal{F}$, if $f\succsim_i g$ for some $i\in N$ and $f(s) \succsim_j g(s)$ for all $s\in S$ and $j\in N \backslash \{i \}$, then $f\succsim_0 g$. 
\end{description}

Liberalism postulates that for any pair of  alternatives that is relevant only for individual $i$ (e.g., the color of shirts that $i$ wears), the planner should respect $i$'s opinion (cf.~\citet{sen1970impossibility}). 
Dominance Pareto considers the case where all individuals except for individual $i$ think an act $f$ is weakly better than another act $g$ in every state.
Then, perception functions of these individuals do not matter, and hence only the perception function of $i$ is relevant. 
This axiom requires that if $i$ also weakly prefers $f$ to $g$ in such a situation, then $f$ should be socially at least as good as $g$.
The following result characterizes the aggregation rules obtained in Theorem 1.%
\footnote{The author is grateful to a referee for suggesting Dominance Pareto.}

\begin{proposition}
\label{prop:iff}
Suppose that for each $i\in N\cup \{ 0 \}$, $\succsim_i$ is  a variational Bewley preference $(u_i, c_i)$ and  Preference Diversity holds. 
Then, the following three statements are equivalent:
\begin{enumerate}[(i)]
    \item Constant Pareto and Liberalism hold; 
    \item Dominance Pareto holds; and
    \item there exists $(\alpha, \beta) \in \mathbb{R}^n_+ \times \mathbb{R}$   such that 
    \begin{equation}
    \label{eq:propiff}
        u_0 = \sum_{i\in N} \alpha_i u_i + \beta, ~~ \text{and} ~~
        c_0 (p) \geq \max_{i\in N} \alpha_i c_i(p) ~~~ \forall p\in \Delta(S).
    \end{equation}
\end{enumerate}
\end{proposition}

\begin{proof}
    We can prove that each of (i) and (ii) implies (iii) in the same way as Theorem \ref{thm:main}. 
    We show that (iii) implies (i). 
    Suppose that \eqref{eq:propiff} holds for some $(\alpha, \beta) \in \mathbb{R}^n_+ \times \mathbb{R}$.
    It is straightforward to check that Constant Pareto holds.
    Take $i\in N$ arbitrarily. Let $f,g\in \mathcal{F}^i$ such that $f\succsim_i g$, that is, $\sum_{s\in S} p(s) u_i (f(s)) + c_i (p) \geq  \sum_{s\in S} p(s) u_i (g(s))$ for all $p\in \Delta(S)$. By the definition of $\mathcal{F}^i$, $u_j(f(s)) = u_j(g(s))$ for all $s\in S$ and $j\in N\setminus \{ i \}$.  
    Therefore, by \eqref{eq:propiff}, 
    \begin{align*}
        \sum_{s\in S} p(s) u_0 (f(s)) + c_0 (p) 
        &\geq \sum_{s\in S} p(s) \bigg(  \sum_{j\in N} \alpha_j u_j (f(s)) + \beta \bigg) + \alpha_i c_i (p) \\
        &= \sum_{s\in S} p(s) \sum_{j\in N\setminus \{ i \}  } \alpha_j u_j (f(s)) + \beta + \alpha_i \bigg(\sum_{s\in S} p(s) u_i (f(s)) + c_i (p)\bigg) \\
        &\geq \sum_{s\in S} p(s) \sum_{j\in N\setminus \{ i \} } \alpha_j u_j (g(s)) + \beta + \alpha_i \sum_{s\in S} p(s) u_i (g(s))  \\
        &= \sum_{s\in S} p(s) u_0 (g(s)). 
    \end{align*}
    Therefore, Liberalism holds. 

    To prove that (iii) implies (ii), suppose that (iii) holds again. Let $f,g\in \mathcal{F}$ be such that  $f\succsim_i g$ for some $i\in N$ and $f(s) \succsim_j g(s)$ for all $s\in S$ and $j\in N \backslash \{i \}$. 
    Then, we have $\sum_{s\in S} p(s) u_i (f(s)) + c_i (p) \geq  \sum_{s\in S} p(s) u_i (g(s))$ and  $u_j(f(s)) \geq u_j(g(s))$ for all $s\in S$ and $j\in N\setminus \{ i \}$.  
    As in the last paragraph, we can show that $\sum_{s\in S} p(s) u_0 (f(s)) + c_0 (p) \geq \sum_{s\in S} p(s) u_0 (g(s))$ for all $p\in \Delta(S)$. 
    Therefore, Dominance Pareto holds.
\end{proof}

Although such an if-and-only-if result holds, we have used Standard Pareto in the main part of this paper because Standard Pareto is a simpler and more basic condition, which can be defined without relying on Preference Diversity. Moreover, starting with Standard Pareto motivates us to conduct the subsequent analysis, such as Propositions \ref{prop:tran} and \ref{prop:iff_ind}. 

\section{Concluding Remarks}\label{sec:con}

We have examined the Paretian aggregation of variational Bewley preferences and showed that under Standard Pareto, the social planner cannot respect all individual preferences in most cases. 
While \citet{danan2016robust} obtained a possibility result using a restricted Pareto principle, 
whether we can obtain such a possibility result by weakening Standard Pareto in a reasonable way remains an open question. 

We finally discuss that the approach taken by \citet{danan2016robust} cannot work well in our framework. 
We say that two acts $f$ and $g$ are \textit{common-taste acts} if $x\succsim_i y$ is equivalent to $x\succsim_j y$ for all $x$ and $y$ in the convex hull of $f(S) \cup g(S)$ and all $i,j\in N$. 
They examined the condition that requires that for all common-taste acts $f,g \in \mathcal{F}$, if $f\succsim_i g$ for all $i\in N$, then $f\succsim_0 g$. 
By imposing the Paretian condition on pairs without taste disagreements, we can separate the perception aggregation from the taste aggregation. 

To conduct a similar analysis, we need to introduce an additional ad-hoc assumption in the above condition to focus on a specific pair of outcomes: for instance, we need to consider the Paretian principles restricted to the pairs of acts on $\{ f\in \mathcal{F} \mid \forall s\in S, ~\exists \alpha \in [0,1]~~\text{s.t.}~~ f(s) = \alpha x^+ + (1-\alpha ) x^- \}$ for some fixed $x^+, x^- \in X$, and the obtained aggregation rules depend on the choice of $x^+$ and $x^-$. 
The key reason behind the difference from \citet{danan2016robust} lies in the uniqueness of parameters in the variational Bewley model. 
In contrast to the Bewley model, the uniqueness of perception functions can
be obtained only when we fix some vNM function in variational Bewley preferences.
Due to this difference, we cannot deal with these parameters separately, which is allowed in the Bewley model.
Therefore, even if we
focus on acts with no disagreement in tastes, we cannot concentrate
on the aggregation of perception functions.

\bibliographystyle{econ}
\bibliography{reference}

\end{document}